\documentclass[aps,tightenlines,a4paper,preprint]{revtex4-2}
\usepackage{graphics,psfrag,amsmath}
\usepackage{epic,eepic}
\usepackage{amsmath}
\usepackage{lmodern}
\usepackage[T1]{fontenc}
\usepackage{textgreek}
\usepackage{color} 
\usepackage{epsfig}
\usepackage{latexsym}
\usepackage{pstricks}
\newcommand{\be}{\begin{equation}}
\newcommand{\ee}{\end{equation}}
\newcommand{\bear}{\begin{eqnarray}}
\newcommand{\eear}{\end{eqnarray}}
\begin{document}

\title{A model for time-evolution of coupling constants}
\author{Taekoon Lee}
\email{tlee@kunsan.ac.kr}
\affiliation{Department of 
Physics, Kunsan National University, Gunsan 54150, Korea}

\begin{abstract}
 A general model is proposed for time-varying
 coupling constants in
  field theory,  assuming the ultraviolet
 cutoff is a varying entity in the expanding universe.
It is assumed that the cutoff depends on
 the scale factor of the  universe and all
   bare couplings remain
  constant. This leads to  varying renormalized coupling 
  constants that evolve in proportion
   to the Hubble parameter. 
   The evolution of the standard model constants is discussed.
\end{abstract}

 

\maketitle  
\newpage

The fundamental laws of physics are generally 
considered universal, independent of time and location.
While this universality is taken for granted  
locally, it is an intriguing question whether it
 extends to the entire history
of the universe.
Noting the similarity in the large ratios, of order $10^{39}$, 
of the electric to gravitational forces
between an electron and a proton and the age of the
universe in atomic time units, Dirac proposed 
that such a large number has to do with the
age of the universe, with the gravitational constant being inversely 
proportional to it \cite{Dirac:1937ti}. While this turned out to be incompatible 
with  various evidences, including the geological ones \cite{Teller:1948zz}, 
the question on the constancy of 
physical constants   over 
a cosmological time scale has been of great interest ever since \cite{Uzan:2010pm,Martins_2017}. 

The  implication of time dependence of the physical laws would 
unquestionably be revolutionary. 
 Above all, varying constants  violate the equivalence 
 principle of general relativity that
the physical laws be same to all free falling observers,
 which would require  a modification 
of the general relativity, and perhaps a fundamental 
change in our understanding of the spacetime. 

Being  accurately measurable the fine structure 
constant $\alpha$ has attracted a special attention in the study of
 time-variation of physical constants, 
and a broad class of theoretical models on varying $\alpha$ is
 based on  scalar fields, which include  the dilaton from 
  the string theory or a Kaluza-Klein mode of higher-dimensional 
  spacetime models.
In these models the scalar field couples to the electromagnetic fields non-minimally,
and  the evolution of the  scalar
   field renders the
fine structure constant vary in time and space. 
In other models based on grand unified theories (GUTs) a varying 
fine structure constant would  imply  varying 
 gauge couplings of the strong, weak interactions as well, resulting in a
 varying nucleon mass \cite{CALMET2002173,PhysRevD.67.015009}.

In this note we propose a  mechanism 
for time-varying coupling constants in field theories in general. 
It may appear that a field theory such as 
the standard model would not
 allow varying coupling constants within its framework.
 There is however a venue: the ultraviolet (UV) cutoff.
 In a renormalizable theory of elementary particles the cutoff is usually
  introduced to regularize the
 divergence of loop amplitudes.  The  cutoff could be an artifact 
 of renormalization process, and
 in asymptotically free theories like the quantum chromodynamics (QCD)
  it needs not have a physical meaning and 
 can be put to an infinite limit. On the other hand in a theory like 
 quantum electrodynamics (QED), which is 
 thought to be a trivial theory in the infinite cutoff 
 limit, the cutoff should be finite
 for the theory to have non-vanishing interactions. 
 In this case the cutoff would be physical
  and the theory must be considered as a low energy effective theory.
In this note we consider a field theory as an effective theory and assume 
the UV cutoff is finite and physical. We further suppose the cutoff is also time-varying.
Since the renormalizable couplings 
are  related to the bare couplings at the cutoff scale via the
 renormalization group equations (RGEs), they
will vary as the cutoff varies, provided the bare couplings remain constant. 

  As an  illustration  let us consider 
the QCD. Because the QCD is asymptotic free and UV safe, 
the cutoff may be rendered to infinity. In the chiral limit, where quarks are massless,
the nucleon mass $m_{n}$ can be written in the form
\bear{}
m_{n}=\Lambda f(g_{\Lambda})
\label{eq1}
\eear{}
where $\Lambda, g_{\Lambda}$ are the UV cutoff and the bare coupling, respectively, 
and $f$ is a function given by the beta function of the coupling.
In renormalization theory, $m_{n}$ is 
invariant under the variation of the cutoff because
the bare coupling varies as well according to the RGE 
so that $m_{n}$ remains constant.
We may now assume, however, the QCD is an 
effective theory built on a lattice, like the lattice QCD,
but the cutoff, the inverse of lattice spacing, is physical, and time-dependent,
  while the bare coupling remains at a fixed but 
  small value so that the theory remains near the
  continuum limit. 
The nucleon mass  will then vary  in proportion to
the cutoff as given in Eq. (\ref{eq1}) --- a varying nucleon mass.

This idea can be easily implemented on a general renormalizable theory.
Let us consider 
  the RGEs for a set of $N$ dimensionless
  couplings $\alpha_i(\mu)$, where $i=1,2,\cdots,N$:
\bear
\mu\frac{d  \alpha_i(\mu)}{d\mu}=
\beta_i(\alpha_1(\mu),\cdots,\alpha_N(\mu)){}\,,
\label{rge}
\eear
where the couplings and the beta functions are from a 
mass-independent renormalization scheme like the MS scheme.
We shall now assume  that 
the  cutoff varies under the 
expansion of the universe, while the
 bare couplings $\alpha_i^{\rm B}\equiv\alpha_i(\Lambda)$
remain constant. To see the variation of $\alpha_{i}(\mu)$ under this
boundary condition we write 
 the solution of Eq. (\ref{rge}) in the form:
\bear
\alpha_i(\mu)=F_i(\alpha_1^{\rm B},\cdots,{}
\alpha_N^{\rm B},\log(\Lambda/\mu))\,,
\nonumber\eear{}
from which we get
\bear
\delta \alpha_i(\mu)&=&\frac{\partial 
F_i}{\partial \log\Lambda}
\frac{\delta \Lambda}{\Lambda} \nonumber \\
&=& -\frac{\partial F_i}{\partial \log\mu}
\frac{\delta \Lambda}{\Lambda}\nonumber \\
&=& -\beta_i(\alpha_1(\mu),\cdots,\alpha_N(\mu)){}
\frac{\delta \Lambda}{\Lambda}\,.
\label{mastereq}
\eear
Note that the variation of 
the couplings at the scale  $\mu$  is local in that it
depends only on the beta function values at the same 
scale, oblivious of the evolution history of the couplings.
In deriving Eq.~(\ref{mastereq}) we assumed there was no 
threshold between the cutoff and $\mu$.{}
In case there is one  Eq.~(\ref{mastereq}) 
can be used to compute the variation of the coupling
at just above the matching scale, then using the 
matching function obtain the coupling just below it,
 and run the coupling down to  obtain the variation at $\mu$.

What causes the cutoff vary? We  cannot answer it but there have been 
speculations that  the spacetime may be
discrete and lattice-like at a short distance. 
If this is the case then the lattice spacing would be the cutoff, 
and it may not be inconceivable that the lattice varies as well,
 as the universe expands, causing the cutoff vary.
   Whatever the cause may be we  assume that 
the cutoff varies in the expanding
universe, and the variation depends on  the scale factor of the
Robertson-Walker metric.  
As a simple ansatz we shall assume a
 power-law variation,
 \bear
\Lambda \propto a(t)^{-\kappa}\,,
\label{powerlaw}
\eear
 where $\kappa$ is a constant and $a(t)$ is the scale factor.
  Since  $\kappa$
  determines the strength of 
  the time-dependence of the cutoff, we may think  of it as
 a measure of the
 rigidity  of the field theory
 in the expanding universe.
 
 With the ansatz we have
 \bear{}
 \dot{\Lambda}/\Lambda=-\kappa H\,,{}
 \label{timerate}
\eear{}
 where $H$ is the Hubble parameter. 
 Then with (\ref{mastereq}) we get
 \bear{}
 \frac{d\alpha_i(\mu)}{dt}
=\kappa \beta_i(\alpha_1(\mu),\cdots,\alpha_N(\mu))H\,,
\label{timedependence}
\eear
 which shows  the sign of variation of a
  coupling constant is dependent on its beta function.
  Thus the fine structure constant and the strong coupling constant 
  of QCD evolve in the opposite direction
  as their beta functions have  opposite signs.
  With Eq. (\ref{timedependence}) the variation of the
 coupling from the epoch of red shift $z$ to {\it now} is given by
\begin{widetext}
 \bear{}
 \Delta\alpha_{i}&\equiv& 
 \alpha_{i}^{z}(\mu)-\alpha_{i}^{0}(\mu)= -\kappa \beta_{i}(\alpha_{1}^{0}
 (\mu),\cdots,\alpha_{N}^{0}(\mu))\log(1+z){} +{\cal O}(\kappa^{2})\,, 
 \label{z-dependence}
 \eear{}
 \end{widetext}
 where $\alpha_i^{0}$ are the couplings at present. So  
the couplings vary logarithmically 
 in time as $\kappa$ is expected to be very small, and 
 for a large $z$,   $1+z$ is proportional to
  $t^{-2/3}$  and $t^{-1/2}$ for the matter dominated and
 radiation dominated epochs, respectively.
 
   In the following we investigate  the implication of the result 
   on the standard model couplings, using 
    the leading-order beta functions.
 An immediate consequence of our scenario  is  that
       the renormalized couplings have
        time-variations that are all
     related, as the variations arise from a single cutoff.

 {\bf The fine structure constant:}
The standard model 
couplings $\alpha_{i}$ ($i=1,2,3$) 
  run at one-loop order  by
\bear{}
\mu\frac{d \alpha_{i}}{d\mu}= b_{i} \alpha_{i}^{2}\,,
 \label{onelooprunning}
 \eear
 where 
$ b_{i}=(41/6,-19/6,-7)/2\pi $ for $\mu>M_{Z}$,
 and $\alpha_{i}=g_{i}^{2}/4\pi$, $g_i$  the gauge couplings of the
  $U(1)_{Y}\times{\rm SU}(2)_{\rm L}\times{\rm SU}(3)_{C}$.
 At the leading order the 
 threshold effects in the running coupling 
 below the electroweak
  symmetry breaking scale can be ignored, and
 the variation of the fine structure constant $\alpha$ can be written
  in terms of those of
 the electroweak couplings at $\mu=M_{Z}$.
 Then we get
 \bear{}
 \delta (1/\alpha)&=& \delta (1/\alpha_{1}(M_{Z}))
  +\delta (1/\alpha_{2}(M_{Z}))
                   =\frac{11}{6\pi}\frac{\delta\Lambda}{\Lambda}\,,
 \nonumber \eear{} 
 and
  \bear{}
  \dot{\alpha}/\alpha&=&\frac{11\kappa}{6\pi}\alpha H\,,
  \quad
  \Delta \alpha/\alpha =-
  \frac{11\kappa}{6\pi}\alpha\log(1+z){}\,.
  \eear{}
  
  The parameter $\kappa$ can be constrained by 
 astronomical observations as well as laboratory experiments.
  The only observation of variation of the
     fine structure constant that is
    significantly different from 
zero is
 by Webb et.al. from 
    the quasar absorption spectra \cite{PhysRevLett.82.884,Murphy:2003hw}, with weighted mean:
    \bear{}
    \Delta\alpha/\alpha=(-0.57\pm0.11)\times 10^{-5}\,,
\quad \text{for}\quad
0.2<z<4.2\,,
    \nonumber\eear{}
    which gives 
    \bear{}
    0.66\times10^{-3}<\kappa<0.88\times 10^{-2}\,.
   \label{const-qsr}
    \eear{}
   There are many other studies that do not  confirm the
    time-variation
    of $\alpha$ \cite{Uzan:2010pm}, and it would be safe to assume that
   \bear{}
   |\kappa|\lesssim10^{-2}\,.
   \nonumber\eear
   
   A laboratory experiment using atomic
    clocks can also give a stringent limit.
     The experiment on the variation of 
    the frequency ratio of
    ${\rm Al}^{+}$ 
    and ${\rm Hg}^{+}$ single ion optical clocks yields a bound \cite{doi:10.1126/science.1154622},
    \bear{}
    \dot{\alpha}/\alpha=(-1.6\pm2.3)\times{}
     10^{-17}/{\rm year}\,,
    \nonumber\eear{}
    which gives
    \bear{}
    \kappa=(-5.4\pm7.8)\times 10^{-5}\,,{}
    \label{const-clock}
    \eear{}
    and  the latest experiment on the frequency ratio of the electro-octupole and 
    electro-quadrupole transitions in ${\rm Yb}^{+}$ yields \cite{PhysRevLett.130.253001}
    \bear{}
    \dot{\alpha}/\alpha=(1.8\pm2.5)\times{}
     10^{-19}/{\rm year}\,,
    \nonumber\eear{}
    which gives
    \bear{}
    \kappa=(6.1\pm8.5)\times 10^{-7}\,.{}
    \label{const-clock1} 
    \eear{}
 
 {\bf The strong coupling constant 
 and the nucleon mass:}
  A similar calculation for the strong coupling constant gives
  \bear{}
   \delta(1/\alpha_{3}(\mu))=-\frac{7}{2\pi}\delta\Lambda/\Lambda \,,\quad
   \dot{\alpha}_{3}/\alpha_{3}=
    -\frac{7\kappa}{2\pi}\alpha_{3} H\,,
  \eear 
  where $\mu$ is of the nucleon mass scale.
   In the chiral limit the nucleon mass $m_{n}$ is
     proportional to $\Lambda_{\rm QCD}$ given by
    \bear{}
    1/\alpha_{3}(\mu)=\beta_{0}
    \log(\mu/\Lambda_{{\rm QCD}})\,,{}
    \label{qcd2}
    \eear{}
    where $\beta_{0}=9/2\pi$ for three light-quark flavors.
    Therefore,
    \bear{}
   \frac{\delta m_{n}}{m_{n}}= \frac{\delta\Lambda_{\rm QCD}}
     {\Lambda_{\rm QCD}}
     =\frac{7}{9}\frac{\delta
    \Lambda}{\Lambda}\,,
    \label{lambdaQCD}
    \eear{}
    which shows the variation of the nucleon mass 
    mirrors that of the cutoff and gives
   \bear{}
   \frac{\dot{m}_{n}}{m_{n}}=
   -\frac{7}{9}\kappa H\,,\quad
   m^{z}_{n}=m^{0}_{n}
   \left[1+\frac{7}{9}\kappa \log (1+z)\right]\,,
   \eear{}
   where $m_n^z$ is the nucleon mass at redshift $z$.
    The ratio of the nucleon mass variation to that of the
   fine structure constant, which has a
    strong model dependence, is given by
   \bear{}
   R=\left(\frac{\dot{m}_n}
   {m_n}\right){}
   \bigg/\left(\frac{\dot{\alpha}}{\alpha}\right)=
   -\frac{14\pi}{33\alpha}=-183\,,
   \nonumber\eear{}
 which compares to  
 the SU(5) GUT model value $R=36$ \cite{CALMET2002173,PhysRevD.67.015009}.
  
  {\bf The lepton, quark, and Higgs masses:}
  The running  mass for these particles is given in the form
  \bear{}
  m(\mu)=\xi(\mu) v(\mu)\,,
  \eear{}
  where $v$ denotes the Higgs vacuum expectation value,
    $\xi$ the Yukawa coupling for
   the lepton and quark mass, and
   $\xi=\sqrt{\lambda}$ for the Higgs mass, where $\lambda$ 
   is the Higgs quartic coupling.
   The RG equation for $m$ is given by
   \bear{}
   \mu\frac{dm(\mu)}{d\mu}=\gamma_m(\mu) m(\mu){}\,,
   \label{RG-mass}
   \eear{}
   where $\gamma_m(\mu)=\gamma_{\xi}(\mu)+
  \gamma_{v}(\mu) $
with
   \bear{}
   \mu\frac{d\xi(\mu)}{d\mu}=\xi(\mu)\gamma_{\xi}(\mu)\,,\quad
  \mu\frac{d v(\mu)}{d\mu}=v(\mu)\gamma_{v}(\mu)\,.
  \nonumber
  \eear{}
 Similarly to the variation of the couplings in (\ref{mastereq}) 
the mass variation is given by
  \bear{}
  \frac{\delta m(\mu)}{m(\mu)}=-\gamma_m(\mu)
 \frac{\delta 
  \Lambda}{\Lambda}\,,\quad \frac{\dot{m}}{m}=\kappa\gamma_mH \,.
  \eear{}
  The anomalous dimension for $v$ can be written as
  \bear{}
  \gamma_{v}=(\gamma_{m_{H}^{2}}-\beta_{\lambda}/\lambda)/2{}\,,
  \eear
  where $\gamma_{m_{H}^{2}}$ is the anomalous dimension of the Higgs mass squared and
  $\beta_{\lambda}$ is the beta function for $\lambda$.
 
 For the electron mass $m_e$
 the anomalous dimension is given by
 \bear
 \gamma_{m_e}= -\frac{3}{16\pi^2}\left[\lambda 
 + g_1^2+\frac{1}{8\lambda}(g_1^4+
 2g_1^2g_2^2 +3 g_2^4-16Y_t^4)\right]\,, 
 \eear
 where $Y_t$ is the top quark Yukawa coupling \cite{Luo:2002ey}.
 Evaluating it at $\mu=M_Z$ with $g_1=0.350,g_2=0.653,Y_t=0.935$, 
 and $\lambda=0.265$ we have
 \bear
 \frac{\delta m_e}{m_e}=-0.096 \frac{\delta\Lambda}{\Lambda}\,,
 \eear
 which is about 1/8th of the nucleon mass variation of the opposite sign.
 
  The variation of the
   proton-to-electron mass ratio, 
   $\zeta=m_p/m_e$, is given by
   \bear{}
    \dot{\zeta}/\zeta=-0.874\kappa H\,, \quad
 \Delta{\zeta}/\zeta= 0.874\kappa \log (1+z)\,.
   \nonumber\eear{}
   The variation for  $\zeta$ 
   can be constrained by molecular 
   transition lines, and Reinhold et al. 
   observed a non-vanishing value
   $\Delta \zeta/\zeta={}
   (2.4\pm0.6)\times 10^{-5}$ 
   from a weighted fit of ${\rm H}_{2}$ spectral lines 
   from two quasars at
   redshift $z=3.02$ and $z=2.59$ \cite{PhysRevLett.96.151101}, which yields 
   $\kappa=(2.4\pm 0.6)\times 10^{-5}$.
   However, the inversion spectrum
   of ammonia yields a tighter constraint of
    $|\Delta \zeta/\zeta|{}
    <1.8\times10^{-6}$ 
    from the absorption lines of quasars at
   redshift $z=0.685$, 
   which gives $\kappa< 4\times 10^{-6}$ \cite{doi:10.1126/science.1156352}.
   The laboratory experiments probing frequency drift of 
   atomic clocks also give stringent constraints on
   the differential variation:
    $\dot{\zeta}/\zeta=(-5.3\pm 6.5)
    \times 10^{-17}/{\rm yr}$ \cite{McGrew:19},
     yielding $\kappa=(0.9\pm1.1)\times 10^{-6}$.
   
  {\bf The gravitational constant $\boldsymbol{G}$:} 
  The Planck scale  
  $M_{\rm pl}=1/\sqrt{G}$ may be a 
   natural candidate for the UV cutoff.
    If  it is the case then
 the gravitational constant varies  as
 \bear{}
 \dot{G}/{G}=
  -2 \dot{\Lambda}/\Lambda=
  2\kappa H\,,\quad
  \Delta{G}/{G}=-2\kappa\log(1+z)\,,
 \eear{}
 with $G$ varying in power law, for example, 
 $G(t)\propto t^{\frac{4}{3}\kappa}$ in the matter dominated epoch.
 A constraint on the power law variation of  $G$  has been 
 placed from helioseismology 
 of the sun that probes its evolution under 
 varying $G$ \cite{1994ApJ...437..870D}, 
 from which we get  $|\kappa|\lesssim 0.1$.
 The big bang nuclearsynthesis also provides a constraint on
  the power law variation \cite{ACCETTA1990146},
 which corresponds to $|\kappa|\lesssim 0.008$.

 In summary we proposed  a model for time varying coupling constants 
 in a renormalizable field theory, assuming the
 UV cutoff is varying in the expanding universe while the bare 
 couplings remain fixed. This renders 
 the  renormalized couplings vary in time per
  the renormalization group equations. 
 The evolution of the couplings is proportional to the Hubble parameter and 
 the beta functions of the couplings.  The
 time rates of  the couplings are not independent but  related  
  by the beta functions,
 as they arise from a single UV cutoff.
 Of the standard model constants the most sensitive to 
   the cutoff evolution is the
   nucleon mass, which evolution mirrors the cutoff.

\begin{acknowledgments}
This work was supported by the National Research Foundation of Korea(NRF)
 grant funded by the Korea government(MSIT) (No.2019R1F1A1063011) and KNU.	
\end{acknowledgments}

\end{document}